\documentclass[preprint,superscriptaddress,double-spaced,floatfix,nofootinbib,longbibliography,12pt]{revtex4-1}
\usepackage{hyphenat}
 \usepackage[latin1]{inputenc}
\usepackage{graphicx,amsmath,bbm,bm,array,subfigure}
\usepackage{appendix}
\usepackage{lipsum}
\usepackage{dsfont}
\usepackage{calrsfs}
\usepackage{lineno}
\usepackage{amsmath,amssymb}
\usepackage[linktocpage=true,colorlinks=true,linkcolor=red,citecolor=red,urlcolor=blue]{hyperref}
\usepackage{MnSymbol}

\newcommand{\nn}{\nonumber}

\def\vec#1{\mathchoice
        {\mbox{\boldmath $#1$}}
        {\mbox{\boldmath $#1$}}
        {\mbox{\boldmath $\scriptstyle #1$}}
        {\mbox{\boldmath $\scriptscriptstyle #1$}}
}
\def \beq{\begin{equation}}
\def \eeq{\end{equation}}
\def \beqa{\begin{eqnarray}}
\def \eeqa{\end{eqnarray}}
\def \pd{\partial}
\def \nn{\nonumber}

\newcommand{\mn}{\mu\nu}

\begin{document}

\title {Correspondence between hydrodynamic frames, transport coefficients, and hydrodynamic modes in relativistic fluids}

\author{Md Hasanujjaman}
\email{hasan@apcrgc.org}
\affiliation{Department of Physics, A. P. C. Roy Government College, Siliguri- 734010, India}

\author{Mahfuzur Rahaman}
\email{mahfuzurrahaman01@gmail.com}
\affiliation{Department of Physics, Maulana Azad College, Kolkata - 700013, India}


\begin{abstract}
The choice of hydrodynamic frame directly influences the numerical values of transport coefficients in relativistic dissipative hydrodynamics. We derive the exact transformation between the Eckart and Landau--Lifshitz frames and show that their thermal conductivities are related by an enthalpy-dependent factor. Using a baryon-rich relativistic fluid described by a Boltzmann nucleon gas equation of state, we find that the Landau--Lifshitz thermal conductivity is suppressed relative to the Eckart conductivity, with the difference increasing with temperature and baryon chemical potential. A linearized analysis of sound propagation demonstrates that the sound attenuation coefficient remains identical in both frames, confirming the frame invariance of physical observables. Our results show that the frame dependence of transport coefficients reflects only the different decomposition of dissipative effects into heat-flow and diffusion currents, while the underlying transport physics remains unchanged. These findings underscore the importance of specifying the hydrodynamic frame when comparing transport coefficients from heavy-ion collisions, lattice QCD, and kinetic-theory calculations.
\end{abstract}

\maketitle
\section{Introduction}
\label{sec1}
Hydrodynamics serves as a remarkably powerful qualitative framework for characterizing the collective
motion of QCD (Quantum Chromodynamics) matter generated during Heavy Ion Collisions (HICs). Among its
most notable achievements is the quantitative reproduction of elliptic flow ($v_{2}$) and related flow
observables ($v_{n}$) within a fluid dynamical
picture~\cite{Rischke:1998fq,Shuryak:2003xe,Stoecker:1986ci}. While ideal fluid dynamics proved
quantitatively reliable for central collisions involving large nuclei $(A \approx 200)$ at mid-rapidity
under peak Relativistic Heavy Ion Collider (RHIC) energies, its accuracy progressively deteriorated for
smaller collision systems, peripheral collisions, regions away from mid-rapidity, and reduced collision
energies~\cite{Heinz:2004ar}. Even at the highest attainable centre-of-mass energies $(\sqrt{s})$, the
shear viscosity-to-entropy ratio is bounded from below at $\eta/s = 1/4\pi$, a limit rooted in
black-hole physics through the AdS/CFT correspondence and widely referred to as the KSS
bound~\cite{Kovtun:2004de}. The application of viscous hydrodynamics to high-energy heavy-ion collisions
has since attracted considerable attention, particularly after a remarkably low value of $\eta/s$ was
extracted from elliptic flow measurements~\cite{Romatschke:2007mq}. Analysis of elliptic flow further
indicates that viscous corrections contribute at a level of no less than $30\%$~\cite{Drescher:2007cd}.
Consequently, incorporating viscous effects yields a more accurate description of the strongly coupled
Quark Gluon Plasma (QGP) produced in HICs. Moreover, if QGP is indeed formed in heavy-ion collisions,
its characterization necessarily involves determining transport coefficients such as thermal conductivity,
bulk viscosity, and shear viscosity.

Ideal hydrodynamics is built upon the assumption of local thermodynamic equilibrium, wherein each fluid
element is treated as spatially uniform --- that is, spatial gradients are assumed to vanish (zeroth
order in the gradient expansion). The fundamental variables in this framework are temperature $(T)$,
chemical potential $(\mu)$, and the fluid four-velocity $(u^{\mu})$. Dissipative hydrodynamics, by
contrast, relaxes the strict requirement of local thermodynamic equilibrium, though the system is still
assumed to remain close to equilibrium. The covariant formulation of dissipative fluid dynamics was
independently developed by Carl Eckart in 1940~\cite{Eckart:1940te} and by Landau and Lifshitz in
1987~\cite{Landau_fluid_mechanics}, together forming what is commonly referred to as Navier-Stokes (NS)
theory. In this approach, the entropy four-current ($S^{\mu}$) is expanded in terms of dissipative
fluxes, retaining contributions that are linear in those quantities (i.e., first order in the gradient
expansion of hydrodynamic fields) --- hence the designation first-order theory. The governing
differential equations in first-order theories are parabolic in character, which leads to violations of
causality: perturbations can propagate at superluminal speeds, rendering the theory dynamically unstable.

It should be noted that signal propagation speeds can remain finite even within parabolic theories,
meaning that first-order relativistic hydrodynamic frameworks cannot be dismissed on causality grounds
alone~\cite{Van:2007pw}. Nevertheless, meaningful progress has recently been made in resolving both the
causality and stability issues. The authors of Ref.~\cite{Bemfica_firstorder} demonstrated that adopting
hydrodynamic variables distinct from the thermodynamic ones used by Eckart and Landau-Lifshitz-
namely $T,\,\mu,\,u^{\mu}$- allows the energy-momentum tensor to yield a causal theory. These
findings apply regardless of whether the theory is coupled to Einstein's equations and have been placed
on rigorous mathematical footing. It was further established that linear perturbations around equilibrium
configurations remain stable. A first-order stable theory accommodating non-vanishing baryon chemical
potential was subsequently developed in Refs.~\cite{Kovtun:2019hdm,Bemfica:2020zjp}. In
Ref.~\cite{Bemfica:2020zjp}, it was additionally shown that coupling to Einstein's equations preserves
causality and strong hyperbolicity, and that the resulting system is well-posed when all dissipative
contributions- shear viscosity, bulk viscosity, and heat flow - are retained. Frame-stabilized
formulations within first-order theory, along with their correspondence to second-order descriptions, are
discussed in Refs.~\cite{ArpanDas:2020fnr,ArpanDas:2020gtq,ArpanDas:2020grz}.

The second-order theory was originally formulated by M\"{u}ller~\cite{Muller:1967zza} and subsequently
extended by Israel and Stewart~\cite{Israel:1976tn,Stewart,Israel:1979wp}, giving rise to what is
collectively known as M\"{u}ller-Israel-Stewart (MIS) theory. A defining feature of second-order
relativistic theories is that the entropy current is quadratic in the dissipative fluxes, incorporating
terms of the form
$q_{\mu}q^{\mu}u^{\alpha},\,\Pi^{2}u^{\alpha},\,\pi_{\mu\nu}\pi^{\mu\nu}u^{\alpha},\,\Pi q^{\mu},\,
\pi_{\mu\nu}q^{\mu}$,
and similar expressions that quantify departures from local equilibrium. While the parabolic nature of
first-order equations renders them generally acausal, the hyperbolic structure of second-order equations
ensures causality. The acausality problem is addressed by introducing a finite relaxation time governing
the response of dissipative currents to gradients in the fluid dynamical variables. In second-order
theories, the dissipative fluxes themselves become independent dynamical variables whose evolution
equations describe their relaxation toward the corresponding Navier-Stokes values. A stability analysis
of MIS theory through the study of small perturbations around equilibrium was carried out in
Refs.~\cite{Hiscock:1983zz,Hiscock:1985zz}, confirming that the theory is causal, stable, and
well-posed. More recently~\cite{Bemfica:2019cop,Bemfica:2020xym}, substantial progress has been achieved
in extending the understanding of causality to the nonlinear regime. In particular,
Ref.~\cite{Bemfica:2020xym} established, for the first time, the necessary conditions for causality to
hold nonlinearly in MIS-type theories with both shear and bulk viscosity at vanishing chemical potential.
New second-order formalisms have also been proposed in which covariance and causality are ensured through
the incorporation of memory effects into the irreversible
currents~\cite{Denicol:2008hb,Denicol:2008ha}. Comprehensive reviews of second-order theory are
available in Refs.~\cite{Muronga:2006zw,Denicol:2008rj,Koide:2006ef}.

In the present work, setting aside the debate over the appropriate order of the theory, we restrict our
attention exclusively to the second-order framework --- specifically MIS hydrodynamics --- as it suffices
for our purposes. An additional consideration in dissipative hydrodynamics concerns the choice of frame.
Since the underlying physics must remain independent of this choice, the definitions of the dissipative
fluxes $\Pi,\,\pi^{\mu\nu},\,q^{\mu}$ carry frame dependence. Furthermore, the relaxation structure of
second-order theory enters through a set of coupling and relaxation coefficients,
$\alpha_{0},\,\alpha_{1},\,\beta_{0},\,\beta_{1},\,\beta_{2}$, which are computed via thermodynamic
integrals within the kinetic theory formalism. Coefficients evaluated in two different frames are related
through a constant translation~\cite{Israel:1979wp}. The definitions of the dissipative fluxes are also
sensitive to the choice of metric signature, whether $g^{\mu\nu} = (1,-1,-1,-1)$ or
$g^{\mu\nu} = (-1,1,1,1)$. The present paper only undertakes the investigation of the implications
of the choice of frame.

The paper is organized as follows. Section~\ref{sec2} reviews the thermodynamic framework and the Eckart and Landau--Lifshitz formulations of relativistic hydrodynamics. In Sec.~\ref{sec3}, we derive the transformations of the dissipative fluxes and transport coefficients, while Sec.~\ref{sec4} discusses the transformation properties of the hydrodynamic variables. Section~\ref{sec5} examines how different choices of heat flux lead to different orders of hydrodynamic equations. In Sec.~\ref{sec6}, we investigate sound propagation and demonstrate the frame invariance of the sound attenuation coefficient. The numerical results for a baryon-rich relativistic fluid are presented in Sec.~\ref{sec7}. Finally, Sec.~\ref{sec8} summarizes our main findings.
\section{Covariant thermodynamics}
\label{sec2}
The covariant description of heat transport traces its origins to Fourier's law, which proposes a direct
and instantaneous proportionality between the heat flux and the local temperature gradient,
\beqa
\vec{q} = -\kappa\,\vec{\nabla} T,
\eeqa
where $\vec{q}$, $\kappa$, and $T$ denote the heat flux vector, thermal conductivity, and local
temperature, respectively. Embedded within this elegant simplicity, however, lies a profound physical
shortcoming: the law implicitly demands that thermal disturbances propagate at infinite speed. In other
words, the moment two bodies at different temperatures are brought into thermal contact, heat responds
instantaneously across the entire medium, which is not only physically unrealistic but also
fundamentally violates the principle of causality.

Carlo Cattaneo first restored the causality by endowing the medium with a finite memory, which is characterized by a finite relaxation time $\tau$ as the modification to Fourier's law~\cite{Cattaneo1958}
\beqa
\vec{q} = -\kappa\,\vec{\nabla} T
\quad \longrightarrow \quad
\vec{q} + \tau\,\dot{\vec{q}} = -\kappa\,\vec{\nabla} T,
\eeqa
where the additional term $\tau\,\dot{\vec{q}}$ encodes the inertial delay with which the heat flux
responds to an imposed temperature gradient. As a result, the heat flux does not respond instantaneously to a temperature gradient; instead, it evolves toward its steady-state value over a characteristic timescale determined by $\tau$. This seemingly modest correction elevates the heat equation from a parabolic diffusion equation to a hyperbolic wave equation, widely known as the \textit{telegrapher's equation}, thereby restoring causality by bounding
the speed of thermal signal propagation to a finite value.

The earliest relativistic extension of the heat conduction equation was proposed by Charles Eckart in 1940~\cite{Eckart:1940te}. His formulation established the framework for what are now known as first-order relativistic theories of dissipation. Within this class of theories, the entropy four-current, denoted by \(S^{\mu}\), is not necessarily aligned with the fluid four-velocity \(u^{\mu}\). Moreover, the entropy current is assumed to depend only on quantities that are linear in the deviations from local thermodynamic equilibrium, such as the heat flux and viscous stresses. The local form of the second law of thermodynamics,
\begin{equation}
\partial_{\mu}S^{\mu}\geq 0\,,
\end{equation}
then leads, in the simplest implementation, to the relativistic generalization of Fourier's law,
\begin{equation}
q^{\mu}=-\kappa \Delta^{\mu\nu}
\left(
\partial_{\nu}T+T\dot{u}_{\nu}
\right),
\end{equation}
where \(q^{\mu}\) denotes the heat-flux four-vector, \(\kappa\) is the thermal conductivity, and \(\Delta^{\mu\nu}\) is the projection operator orthogonal to the fluid four-velocity, $u^{\mu}$.

Before proceeding further, it is useful to recall several thermodynamic identities that follow from the first, second, and third laws of thermodynamics. These relations may be written as
\begin{equation}
T\,s=\varepsilon+P-\mu \,n,
\label{eq4}
\end{equation}
where \(T\), \(\varepsilon\), \(P\), \(\mu\), \(n\), and \(s\) denote the temperature, energy density, pressure, chemical potential, particle number density, and entropy density, respectively. The Gibbs--Duhem relation takes the form
\begin{equation}
dp=s\,dT+n\,d\mu\,,
\label{eq5}
\end{equation}
from which one obtains
\begin{equation}
T\,ds=d\varepsilon-\mu\,dn\,.
\label{eq6}
\end{equation}

To cast these relations into a covariant form, we introduce the quantities
\begin{equation}
\beta=\frac{1}{T}, \qquad
\beta_{\mu}=\frac{u_{\mu}}{T}, \qquad
\alpha=\frac{\mu}{T}\,.
\end{equation}
Using these definitions, Eqs.~(\ref{eq4})--(\ref{eq6}) can be postulated covariantly as~\cite{Israel:1979wp}
\begin{equation}
S^{\mu}_{(0)}
=
\beta_{\nu}T^{\mu\nu}_{(0)}+P\beta^{\mu}
-\alpha N^{\mu}_{(0)}\,,
\label{eq8}
\end{equation}
\begin{equation}
dS^{\mu}
=
-\alpha\, dN^{\mu}
+\beta_{\nu}\, dT^{\mu\nu}\,,
\label{eq9}
\end{equation}
\begin{equation}
d\!\left(p\beta^{\mu}\right)
=
N^{\mu}_{(0)}\, d\alpha
+
T^{\mu\nu}_{(0)}\, d\beta_{\nu}\,,
\label{eq10}
\end{equation}
where \(T^{\mu\nu}\) and \(N^{\mu}\) represent the energy--momentum tensor and particle four-current, respectively, while \(T^{\mu\nu}_{(0)}\) and \(N^{\mu}_{(0)}\) correspond to their equilibrium values in ideal hydrodynamics.

The extension from equilibrium to non-equilibrium states is based on the assumption that Eq.~(\ref{eq10}) remains valid for arbitrary deviations rather than being restricted to infinitesimal departures from equilibrium. Consequently, combining Eqs.~(\ref{eq8}) and (\ref{eq9}) yields
\begin{equation}
S^{\mu}_{(0)}+dS^{\mu}
=
P\beta^{\mu}
-\alpha\left(N^{\mu}_{(0)}+dN^{\mu}\right)
+\beta_{\nu}\left(T^{\mu\nu}_{(0)}+dT^{\mu\nu}\right)\,,
\label{eq11}
\end{equation}
which describes the entropy current of an arbitrary non-equilibrium state.

When higher-order deviations from local equilibrium are included, the entropy four-current acquires additional contributions and may be written as
\begin{equation}
S^{\mu}
=
P\beta^{\mu}
-\alpha N^{\mu}
+\beta_{\nu}T^{\mu\nu}
-\mathcal{Q}^{\mu}\,,
\label{eq12}
\end{equation}
where \(\mathcal{Q}^{\mu}\) is a function of the departures from equilibrium, and one can define:
\begin{equation}
\delta T^{\mu\nu}
=
T^{\mu\nu}-T^{\mu\nu}_{(0)},
\qquad
\delta N^{\mu}
=
N^{\mu}-N^{\mu}_{(0)}\,.
\end{equation}
The quantity \(\mathcal{Q}^{\mu}\) may be expanded as a Taylor series in the dissipative fluxes. Retaining terms up to second order in these deviations gives rise to the second-order theory of relativistic dissipative hydrodynamics. The specific form of \(\mathcal{Q}^{\mu}\) determines the constitutive relations and, consequently, the evolution equations governing the dissipative currents.

\section{M\"{u}ller-Israel-Stewart Hydrodynamics}
\label{sec3}
The covariant formulation of relativistic dissipative hydrodynamics was pioneered independently by Eckart~\cite{Eckart:1940te}, and by Landau and Lifshitz~\cite{Landau_fluid_mechanics}. These formulations belong to the class of first-order theories, wherein the dissipative fluxes are assumed to respond instantaneously to thermodynamic forces. While providing a natural relativistic extension of classical irreversible thermodynamics, these theories suffer from fundamental shortcomings, including acausality and the instability of equilibrium states~\cite{Hiscock:1983zz,Hiscock:1985zz}.

In an effort to resolve these deficiencies, M\"uller introduced a second-order theory of relativistic irreversible thermodynamics in 1967 \cite{Muller:1967zza}. This framework was subsequently refined and systematically developed by Israel and Stewart, culminating in the formulation of the Israel--Stewart (MIS) theory of relativistic dissipative hydrodynamics~\cite{Israel:1976tn,Stewart,Israel:1979wp}. By incorporating second-order deviations from local equilibrium into the entropy current, the MIS theory gives rise to hyperbolic transport equations for the dissipative fluxes and naturally introduces finite relaxation times. The causal and stability properties of the MIS framework were later investigated by Hiscock and Lindblom~\cite{Hiscock:1983zz,Hiscock:1985zz}. Their analysis demonstrated that, in contrast to the first-order theories, the MIS formulation remains causal and stable over a considerably wider range of physical conditions. Therefore, MIS theory provides a consistent and robust theoretical foundation for the description of relativistic dissipative fluids.

\subsection{The equations of motion}
The M\"uller-Israel-Stewart (MIS) theory describes a relativistic dissipative fluid in terms of the energy-momentum tensor \(T^{\mu\nu}\) and the particle four-current \(N^\mu\). Adopting the metric convention \(g^{\mu\nu}=(-1,+1,+1,+1)\), the energy--momentum tensor may be decomposed into an equilibrium contribution and a dissipative correction as
\begin{equation}
T^{\mu\nu}=T^{\mu\nu}_{(0)}+\tau^{\mu\nu}.
\end{equation}
For a fluid in local thermodynamic equilibrium, the energy-momentum tensor takes the ideal-fluid form
\begin{equation}
T^{\mu\nu}_{(0)}
=
\varepsilon u^\mu u^\nu
+
P\Delta^{\mu\nu},
\end{equation}
where \(\varepsilon\) and \(P\) denote the local energy density and equilibrium pressure, respectively. In the presence of dissipative processes, the energy--momentum tensor acquires additional contributions and can be expressed as
\begin{equation}
T^{\mu\nu}
=
\varepsilon u^\mu u^\nu
+
(P+\Pi)\Delta^{\mu\nu}
+
h^\mu u^\nu
+
h^\nu u^\mu
+
\pi^{\mu\nu},
\label{tmunu}
\end{equation}
where \(\Pi\) represents the bulk viscous pressure, \(h^\mu\) denotes the energy-diffusion (or heat-flow) four-vector, and \(\pi^{\mu\nu}\) is the shear-stress tensor. These quantities characterize deviations from local equilibrium and encode the dissipative properties of the fluid.

Similarly, the particle four-current may be decomposed into equilibrium and dissipative parts according to
\begin{equation}
N^\mu=N^\mu_{(0)}+\nu^\mu,
\end{equation}
where the equilibrium contribution is given by
\begin{equation}
N^\mu_{(0)}=nu^\mu.
\end{equation}
Consequently, the full particle current assumes the form
\begin{equation}
N^\mu=nu^\mu+\nu^\mu,
\label{nmu}
\end{equation}
where \(n\) is the particle number density (baryon, charge, strangeness) measured in the local rest frame and \(\nu^\mu\) denotes the particle-diffusion current. The dissipative currents \(h^\mu\) and \(\nu^\mu\) are orthogonal to the fluid four-velocity and therefore describe transport processes occurring in the spatial hypersurface perpendicular to \(u^\mu\). Together, the bulk viscous pressure, heat flow, diffusion current, and shear-stress tensor constitute the nonequilibrium degrees of freedom that distinguish the MIS theory from ideal relativistic hydrodynamics. The particle diffusion current is related to the heat flow current $h^{\mu}$ as
\beqa
q^{\mu}=h^{\mu}- \frac{\varepsilon+P}{n}\,\nu^{\mu}\,,
\label{eq21}
\eeqa
where  $q^{\mu}$ is termed as heat flow vector, the $u^{\mu}=\gamma\,(1,\vec{v})$ is the fluid four velocity, and follows
\beqa
u_{\mu}u^{\mu}=-1, 
\label{eq18}
\eeqa
Here $\gamma=1/(1-v^{2})$, is the Lorentz factor.  $\Delta^{\mn}$ is the orthogonal projection operator to $u^{\mu}$, and is defined as
\beqa
\Delta^{\mn}=g^{\mn}+u^{\mu}u^{\nu}\,.
\eeqa
 It has the properties 
 \beqa
 \Delta^{\mu\nu}u_{\mu}=\Delta^{\mu\nu}u_{\nu}=0, \,\,\,\,\,  \,\,\,\, \Delta^{\mu\nu}\Delta^{\alpha}_{\nu}=\Delta^{\mu\alpha}, \,\,\,\, \Delta^{\mn}\pd_{\nu}=\nabla^{\mu},\,\,\,\,\Delta^{\mu}_{\mu}=3\,.
 \eeqa
The heat flux follows the orthogonality conditions as:
\beqa
u_{\mu}h^{\mu}=0, \,\,\,\, u_{\mu}\nu^{\mu}=0\,.
\label{eq19}
\eeqa
 The viscous fluxes also follows the orthogonality as the following:
\beqa
\label{eq22}
u_{\mu}q^{\mu}&=& q^{\mu}u_{\mu}=0\,.
\eeqa
As the shear stress tensor is a two-rank symmetric traceless tensor, it has the properties:
\beqa
u_{\mu}\pi^{\mu\nu}=0,\,\,\,\, \pi^{\mn}&=&\pi^{\nu\mu},\,\,\,\,
\pi^{\mu}_{\mu}=0\,.
\label{eq23}
\eeqa
The relations between energy density, pressure and the dissipative fluxes and EMT are given by the following relations:
\beqa
q_{\alpha}&=&u_{\mu}\tau^{\mu\nu}\Delta _{\nu\alpha}, \,\,\,\, u_{\mu}\tau^{\mu\nu}=q^{\nu}\,, \nn\\
P+\Pi&=&-\frac{1}{3}\Delta_{\mu\nu}T^{\mu\nu}, \,\,\,\, \Pi=-\frac{1}{3}\Delta_{\mu\nu}\tau^{\mu\nu}
\eeqa
The equations of motion will then be
\beqa
\label{eq25}
\pd_{\mu}T^{\mn}=0\,,\\
\pd_{\mu}N^{\mu}=0\,.
\label{eq26}
\eeqa
The energy--momentum tensor \(T^{\mu\nu}\), defined in Eq.~\eqref{tmunu}, is a symmetric rank-two tensor and therefore possesses ten ($10$) independent components. Similarly, the particle four-current \(N^{\mu}\), introduced in Eq.~\eqref{nmu}, contributes four ($4$) additional independent components, resulting in a total of fourteen independent hydrodynamic variables. In contrast, the decomposition of \(T^{\mu\nu}\) and \(N^{\mu}\) according to Eq.~\eqref{eq19} introduces the dissipative currents \(q^{\mu}\) and \(\nu^{\mu}\), each orthogonal to the fluid four-velocity and hence characterized by three ($3$) independent components. Furthermore, the shear-stress tensor \(\pi^{\mu\nu}\) is symmetric, traceless, and orthogonal to \(u^{\mu}\), leaving five ($5$) independent components.

On the other hand, the decomposition of \(T^{\mu\nu}\) and \(N^\mu\) introduces the following set of hydrodynamic variables: $\varepsilon\,(1),\,n\,(1),\,P\,(1),\,u^{\mu}\,(4),\,q^{\mu}\,(4),\,\nu^{\mu}\,(4)$, and $\pi^{\mu\nu}\,(6)$ resulting in twenty one $(21)$ variables. But the constraints used in Eqs. \eqref{eq18}, \eqref{eq19}, \eqref{eq22}, and \eqref{eq23} reduce four ($4$) independent components. Consequently, the total number of independent variables in the decomposition becomes $(17)$. This exceeds the original fourteen ($14$) independent components contained in \(T^{\mu\nu}\) and \(N^\mu\). The excess degrees of freedom originate from the arbitrariness in the definition of the fluid four-velocity \(u^\mu\). To remove this ambiguity, three ($3$) additional constraints must be imposed through a suitable choice of hydrodynamic frame. Common choices include the Eckart frame, defined by \(\nu^\mu=0\), and the Landau--Lifshitz frame, for which \(h^\mu=0\). These frame conditions eliminate the redundant degrees of freedom and restore the correct count of fourteen independent variables.

\subsection{Matching conditions, fluid velocity, and the choice of hydrodynamic frame}
In ideal relativistic hydrodynamics, the fluid four-velocity is uniquely defined since the energy flow and the conserved charge flow are parallel to one another. Consequently, the local rest frame (LRF) of the fluid is unambiguously determined, and the energy density and particle number density are identified through the projections
\begin{equation}
\varepsilon=u_{\mu}u_{\nu}T^{\mu\nu},
\qquad
n=u_{\mu}N^{\mu}.
\end{equation}
In this limit, the fluid velocity simultaneously characterizes the transport of energy and conserved charge.

The situation becomes considerably more subtle in the presence of dissipative processes. Heat conduction, particle diffusion, and viscous effects generally cause the energy flux and particle flux to deviate from one another, thereby removing the unique correspondence between the fluid velocity and the underlying transport currents. As a result, the definition of the fluid four-velocity is no longer unique, and an additional prescription is required to specify the local rest frame of the fluid.

To ensure that the nonequilibrium corrections do not modify the locally measured equilibrium energy and particle densities, one imposes the matching (or fitting) conditions
\begin{equation}
u_{\mu}u_{\nu}\tau^{\mu\nu}=0,
\qquad
u_{\mu}\nu^{\mu}=0.
\end{equation}
These conditions guarantee that the quantities \(\varepsilon\) and \(n\) retain their equilibrium interpretations even in the presence of dissipative corrections. Physically, they require that the nonequilibrium terms contribute only to transport processes and not to the definitions of the local thermodynamic variables themselves.

At this stage, however, the four-velocity \(u^{\mu}\) remains an arbitrary normalized timelike vector satisfying $u^{\mu}u_{\mu}=-1$.
To endow \(u^{\mu}\) with a definite physical meaning, it must be related to the conserved currents of the system through an appropriate choice of hydrodynamic frame. Since different conserved quantities may define different local rest frames, several frame choices are, in principle, possible. Among these, the two most commonly employed prescriptions are the Eckart frame~\cite{Eckart:1940te} and the Landau--Lifshitz (LL) frame~\cite{Landau_fluid_mechanics}.

Both frame choices provide a complete and equivalent description of relativistic dissipative fluids. Although physical observables are independent of the chosen frame, the mathematical form of the hydrodynamic equations and the interpretation of the dissipative currents depend on the adopted convention.
\subsection{Eckart frame of reference}
In the Eckart frame, the fluid velocity is defined to be parallel to the conserved particle current,
\begin{equation}
N^{\mu}=nu^{\mu},
\end{equation}
which implies that the particle-diffusion current vanishes. Consequently, the local rest frame is identified with the frame in which the net particle flow disappears, while dissipative effects are entirely encoded in the energy-flow (or heat-flow) vector $h^{\mu}$. Collectively,
\beqa
\nu^{\mu}=0,\,\,\,\,\text{and}\,\,\,\,h^{\mu}\ne 0\,.
\eeqa
The velocity of fluid is defined as
\beqa
u^{\mu}_{E}=\frac{N^{\mu}}{\sqrt{-N^{\nu}N^{\nu}}},\,\,\,\,n=-u^{E}_{\mu}N^{\mu}\,.
\eeqa
 The EMT and the particle current will become
 \beqa
 \label{eq38}
 T^{\mn}_{E}&=&\varepsilon u^{\mu}u^{\nu}+(P+\Pi)\Delta^{\mu \nu}+q^{\mu}u^{\nu}+q^{\nu}u^{\mu}+\pi^{\mu \nu}\,, \\
 N^{\mu}_{E}&=&nu^{\mu}\,.
 \eeqa
 The $14$ unknowns are $\varepsilon,\, n,\, P,\, q^{\mu},\, \pi^{\mn}$, and $u^{\mu}_{E}$.

 \subsection{Landau-Lifshitz frame of reference}
 In contrast, the Landau--Lifshitz frame defines the fluid velocity as the timelike eigenvector of the energy--momentum tensor,
\begin{equation}
T^{\mu\nu}u_{\nu}=-\varepsilon u^{\mu},
\end{equation}
so that the energy-diffusion current vanishes. 
In this frame, the local rest frame corresponds to the frame in which there is no net energy transport, and dissipative effects associated with heat conduction are absorbed into the diffusion current $\nu^{\mu}$. Collectively,
 \beqa
 h^{\mu}=0\,\,\,\,\text{and}\,\,\,\,\nu^{\mu}\ne0\,.
 \eeqa
 The velocity of the fluid is defined as
 \beqa
 u^{\mu}_{L}=-\frac{u_{\nu}T^{\mn}}{\sqrt{u_{\alpha}T^{\alpha \beta}T_{\beta\gamma}u^{\gamma}}},\,\,\,\, u_{\mu}u_{\nu}T^{\mu\nu}=\varepsilon\,.
 \eeqa
 The EMT and the particle current will become
 \beqa
 T^{\mn}_{L}&=&\varepsilon u^{\mu}u^{\nu}+(P+\Pi)\Delta^{\mu \nu}+\pi^{\mu \nu}\,, \\
 N^{\mu}_{L}&=&nu^{\mu}+\nu^{\mu}=nu^{\mu}-\frac{nq^{\mu}}{\varepsilon+P}\,.
 \eeqa
 The $14$ unknowns are $\varepsilon,\, n,\, P,\, \nu^{\mu},\, \pi^{\mn}$, and $u^{\mu}_{L}$.

The above definitions of the fluid four-velocity \(u^{\mu}\) in the Eckart and Landau--Lifshitz (LL) frames impose specific constraints on the dissipative currents and thereby determine how nonequilibrium effects are partitioned among the hydrodynamic variables. In the Eckart frame, the fluid velocity is chosen to be parallel to the conserved particle current. As a consequence, the particle-diffusion current vanishes, and there is no net particle transport in the local rest frame. Any dissipative transport is therefore attributed entirely to the energy (or heat) flow.

In contrast, the LL frame defines the fluid four-velocity through the energy flow to ensure that there is no net energy transport in the local rest frame. In this case, dissipative effects associated with heat conduction manifest themselves through the particle-diffusion current rather than an explicit energy-flow term.

From a physical perspective, the two frame choices correspond to different definitions of what it means for the fluid to be `at rest'. In the Eckart frame, the fluid is at rest with respect to the conserved particles, whereas in the LL frame it is at rest with respect to the flow of energy. Although the decomposition of the dissipative currents differs between the two frames, both descriptions are physically equivalent and lead to the same observable when derived consistently.

\section{Correspondence Between the Eckart and Landau-Lifshitz Frames}
\label{sec4}
The \textit{Eckart frame} is defined by $\nu_E^\mu = 0$ (no particle diffusion), while the \textit{Landau--Lifshitz (LL) frame} is defined by $q_L^\mu = 0$ (no energy diffusion). The difference in hydrodynamic velocities between the frames is proportional to dissipative currents but in equilibrium (no temperature gradients, no chemical potential gradients), both frames coincide as there is no heat flux and no particle diffusion.  The relationship between hydrodynamic velocities are often formalized using a Lorentz transformation as  both $u^{\mu}_{E}$ and $u^{\mu}_{L}$ are unit timelike vectors.  In near-equilibrium (dissipative effects are small or first order in gradient of the dissipative fluxes) they differ only by a small boost \cite{Fotakis:2022usk,Monnai:2019jkc},
\begin{eqnarray}
u_L^\mu = u_E^\mu + \delta u^\mu, \qquad 
u_{E\mu}\delta u^\mu = 0, \label{eq:boost}
\end{eqnarray}
where  $\delta u^\mu$
  is a vector that represents the difference between the frames and is orthogonal to the four-velocity (i.e., it is a purely spatial vector in the local rest frame). This vector $\delta u^\mu$
  is directly proportional to the dissipative currents (like heat flux or diffusion) that distinguish the two frames.
By this choice of transformation, different thermodynamic quantities are transformed as follows
\subsection{Energy density and pressure}
The energy density measured in a frame with velocity $u^\mu$ is $\varepsilon(u) = u_\mu u_\nu T^{\mu\nu}$. Evaluating $\varepsilon_L \equiv \varepsilon(u_L)$:
\begin{eqnarray}
\varepsilon_L &=& (u_{E\mu}+\delta u_\mu)(u_{E\nu}+\delta u_\nu) T^{\mu\nu} \nonumber\\
&=& u_{E\mu}u_{E\nu}T^{\mu\nu} + 2\,\delta u_\mu\, u_{E\nu}T^{\mu\nu} + \delta u_\mu\delta u_\nu T^{\mu\nu}. \label{eq:eps_expand}
\end{eqnarray}
The first term is $\varepsilon_E \equiv \varepsilon(u_E)$. Using the decomposition of $T^{\mu\nu}$ from \eqref{eq38} in the Eckart frame, we obtain:
\begin{eqnarray}
\label{eq45}
u_{E\nu}T^{\mu\nu} = -\varepsilon_E u_E^\mu - q_E^\mu. \label{eq:uT}
\end{eqnarray}
Substituting Eq.\eqref{eq45} into the linear second term of (\ref{eq:eps_expand}):
\begin{eqnarray}
2\,\delta u_\mu (-\varepsilon_E u_E^\mu - q_E^\mu) = -2\varepsilon_E (\delta u_\mu u_E^\mu) - 2\,\delta u_\mu q_E^\mu.
\end{eqnarray}
Orthogonality $\delta u_\mu u_E^\mu =0$ eliminates the first part. With $ q_{E\mu}$ as a quantity of first order in dissipation, one obtains the term $\delta u_\mu q_E^\mu$ as second order in dissipation, and thus neglected. The quadratic term $\delta u_\mu\delta u_\nu T^{\mu\nu}$ in Eq. \eqref{eq:eps_expand} is also second order. Hence, we find
\begin{eqnarray}
\varepsilon_L = \varepsilon_E + \mathcal{O}(\text{diss}^2). \label{eq:eps_order}
\end{eqnarray}
Because the equation of state $P=P(\varepsilon,n)$ is analytic, the pressure satisfies 
\beqa
P_L = P_E + \mathcal{O}(\text{diss}^2).
\eeqa

\subsection{Number density}
The number density in a given frame is $n(u) = -u_\mu N^\mu$. In the Eckart frame, $\nu_E^\mu=0$ so $N^\mu = n_E u_E^\mu$. Then
\begin{eqnarray}
n_L &= -u_{L\mu} N^\mu = -(u_{E\mu}+\delta u_\mu)(n_E u_E^\mu) \nonumber\\
&= -n_E(u_{E\mu}u_E^\mu) - n_E(\delta u_\mu u_E^\mu).
\end{eqnarray}
Using $u_{E\mu}u_E^\mu = -1$ and $\delta u_\mu u_E^\mu = 0$,
\begin{eqnarray}
n_L = n_E. \label{eq:n_exact}
\end{eqnarray}
Thus the number density is exactly the same in the two frames, not merely to second order.
The difference in behavior between the energy density and the number density under a change of hydrodynamic frame can be understood geometrically. The energy density $\varepsilon(u)=u_\mu u_\nu T^{\mu\nu}$ is a quadratic form in the fluid four-velocity $u^\mu$. In the Landau-Lifshitz (LL) frame, the velocity $u_L^\mu$ is defined as the timelike eigenvector of the energy-momentum tensor, satisfying $u_{L\nu}T^{\mu\nu}=-\varepsilon_L u_L^\mu$. Consequently, $\varepsilon(u)$ attains an extremum at $u=u_L$ under the normalization constraint $u_\mu u^\mu=-1$. As shown in Eq.\eqref{eq:eps_expand}, first-order variation of $u$ around this eigenvector produces no first-order change in $\varepsilon$, and thus appears quadratically in the deviation $\delta u$, which itself is of first order in the dissipative fluxes. This property explains why $\varepsilon$, and hence the pressure $P$, through the equation of state differs between the Eckart and LL frames only at second order in the dissipative quantities, i.e., $\varepsilon_L=\varepsilon_E+\mathcal{O}(\text{diss}^2)$.

In contrast, the number density $n(u)=-u_\mu N^\mu$ exhibits a fundamentally different behavior. In the Eckart frame, the particle diffusion current vanishes by construction, so that $N^\mu=n_E u_E^\mu$. When we evaluate $n_L=-u_{L\mu}N^\mu$ using $u_L^\mu=u_E^\mu+\delta u^\mu$, the linear term involves $\delta u_\mu u_E^\mu$, which vanishes identically because $\delta u^\mu$ is orthogonal to $u_E^\mu$. Since $N^\mu$ contains no transverse component in the Eckart frame, the contraction yields $n_L=n_E$ exactly, without any order-by-order approximation. Thus, while the energy density is sensitive to the frame choice only at second order due to the extremal nature of the energy eigenvalue, the number density is strictly frame-invariant in this particular case because the particle current has no dissipative transverse part in the Eckart frame.


\subsection{Four-velocity}

To evaluate $\delta u$, we first compute the term
\beqa
u_{L\nu}T^{\mu\nu}
=
(u_{E\nu}+\delta u_\nu)T^{\mu\nu}.
\eeqa

From $T^{\mn}_{E}=\varepsilon u^{\mu}u^{\nu}+(P+\Pi)\Delta^{\mu \nu}+q^{\mu}u^{\nu}+q^{\nu}u^{\mu}+\pi^{\mu \nu}$, we compute term by term. 

(i) For the ideal energy-density term,
\beqa
(u_{E\nu}+\delta u_\nu)
\varepsilon_Eu_E^\mu u_E^\nu
&=&
\varepsilon_Eu_E^\mu(u_{E\nu}u_E^\nu)
+
\varepsilon_Eu_E^\mu(\delta u_\nu u_E^\nu)
\nn\\
&=&
-\varepsilon_Eu_E^\mu.
\label{eq:T1}
\eeqa

(ii) For the pressure term,
\beqa
(u_{E\nu}+\delta u_\nu)
(P+\Pi_E)\Delta_E^{\mu\nu}
&=&
(P+\Pi_E)\Delta_E^{\mu\nu}u_{E\nu}
\nn\\
&&+
(P+\Pi_E)\Delta_E^{\mu\nu}\delta u_\nu
\nn\\
&=&
P\,\delta u^\mu
+
{\cal O}({\rm diss}^2).
\label{eq:T2}
\eeqa

(iii) For the heat-flow term \(q_E^\mu u_E^\nu\),
\beqa
(u_{E\nu}+\delta u_\nu)
q_E^\mu u_E^\nu
&=&
q_E^\mu(u_{E\nu}u_E^\nu)
+
q_E^\mu(\delta u_\nu u_E^\nu)
\nn\\
&=&
-q_E^\mu .
\label{eq:T3}
\eeqa

(iv) For the term \(q_E^\nu u_E^\mu\),
\beqa
(u_{E\nu}+\delta u_\nu)
q_E^\nu u_E^\mu
&=&
u_E^\mu(q_E^\nu u_{E\nu})
+
u_E^\mu(q_E^\nu\delta u_\nu)
\nn\\
&=&
{\cal O}({\rm diss}^2).
\label{eq:T4}
\eeqa

(v) For the term $\pi^{\mu\nu}$,
\beqa
(u_{E\nu}+\delta u_\nu)
\pi_E^{\mu\nu}
&=&
\pi_E^{\mu\nu}u_{E\nu}
+
\pi_E^{\mu\nu}\delta u_\nu
\nn\\
&=&
{\cal O}({\rm diss}^2).
\label{eq:T5}
\eeqa

Combining Eqs.~(\ref{eq:T1})--(\ref{eq:T5}), finally we obtain
\beqa
u_{L\nu}T^{\mu\nu}
&=&
-\varepsilon_Eu_E^\mu
-q_E^\mu
+P\,\delta u^\mu
+
{\cal O}({\rm diss}^2).
\label{eq:uL_T}
\eeqa
Using the Landau--Lifshitz definition of energy density
\beqa
u_{L\mu}u_{L\nu}T^{\mu\nu}=\varepsilon_{L}
\eeqa
The above can be written as
\beqa
u_{L\nu}T^{\mu\nu}
&=&
-\varepsilon_Lu_L^\mu\nn\\
&=&-\varepsilon_L(u_E^\mu+\delta u^\mu)
\eeqa
Using Eq.\eqref{eq:eps_order}, i.e., $\varepsilon_L=\varepsilon_E+{\cal O}({\rm diss}^2)$,
we obtain
\beqa
u_{L\nu}T^{\mu\nu}
&=& -\varepsilon_L(u_E^\mu+\delta u^\mu)\nn\\
&=&
-\varepsilon_Eu_E^\mu
-\varepsilon_E\delta u^\mu
+
{\cal O}({\rm diss}^2).
\label{eq:eps_L_exp}
\eeqa

Equating Eqs.~(\ref{eq:uL_T}) and (\ref{eq:eps_L_exp}) gives
\beqa
\delta u^\mu
=
\frac{q_E^\mu}
{\varepsilon_E+p}.
\eeqa

Substituting into Eq.\eqref{eq:boost}, we have
\beqa
u_L^\mu
=
u_E^\mu
+
\frac{q_E^\mu}
{\varepsilon+p}
+
{\cal O}({\rm diss}^2).
\label{eq:final_velocity}
\eeqa

Using the relation from Eq.\eqref{eq21},
\beqa
q_E^\mu
=
-\frac{\varepsilon+p}{n}\,
\nu_L^\mu,
\label{eq65}
\eeqa
the inverse transformation becomes
\beqa
u_E^\mu
=
u_L^\mu
+
\frac{\nu_L^\mu}{n}
+
{\cal O}({\rm diss}^2).
\eeqa
\subsection{Transport coefficients}
The correspondence between the Eckart's and LL's frame is established for the energy density, number density, pressure, and the fluid's four velocity. We now examine how the associated transport coefficients transform under a change these frames. Since the equilibrium thermodynamic quantities differ between the two frames only at second order in the dissipative fluxes, the common variables \(\varepsilon\), \(p\), \(n\), and \(T\) may be used throughout the present derivation.

In the Eckart's frame, the heat-flow vector satisfies the relativistic generalization of Fourier's law,
\beqa
q_E^\mu
&=&
-\kappa_E F_E^\mu,
\label{eq:Eckart_law}
\eeqa
where
\beqa
F_E^\mu
\equiv
\nabla^\mu T
+
T\dot{u}_E^\mu,\,\,\,
\nabla^\mu
\equiv
\Delta^{\mu\nu}\partial_\nu,
\,\,\,
\dot{u}^\mu
\equiv
u^\nu\partial_\nu u^\mu .
\eeqa
Here \(\kappa_E\) denotes the thermal conductivity in the Eckart frame. In the LL frame, the dissipative current is represented by the particle-diffusion current,
\beqa
\nu_L^\mu
&=&
-\kappa_L F_L^\mu,
\label{eq:LL_law}
\eeqa
with
\beqa
F_L^\mu
\equiv
\nabla^\mu\alpha
\eeqa
where \(\kappa_L\) is the corresponding diffusion coefficient. The thermodynamic forces appearing in the two frame choices are not independent. To establish the relation, we use the Euler's equation and Gibbs--Duhem relation from Eq.\eqref{eq4} and \eqref{eq5} respectively, along with the relativistic Euler equation,
\beqa
(\varepsilon+p)\dot{u}^\mu
=
-\nabla^\mu p
\eeqa
Now, from the force in LL frame, we start with
\beqa
F_L^\mu&=&\nabla^\mu
\left(\alpha\right)
\nn\\
&=&
\nabla^\mu
\left(
\frac{\mu}{T}
\right)
\nn\\
&=&
\frac{1}{T}\nabla^\mu\mu
-
\frac{\mu}{T^2}\nabla^\mu T
\eeqa
Using Gibbs-Duhem relation, we have
\beqa
\nabla^\mu\mu
=
\frac{1}{n}\nabla^\mu p
-
\frac{s}{n}\nabla^\mu T,
\eeqa
and
\beqa
\dot{u}^\mu
=
-\frac{1}{\varepsilon+p}
\nabla^\mu p,
\eeqa
we obtain
\beqa
F_L^\mu
&=&
\frac{1}{nT}\nabla^\mu p
-
\frac{s}{nT}\nabla^\mu T
-
\frac{\mu}{T^2}\nabla^\mu T
\eeqa
The above expression simplifies to
\beqa
F_L^\mu
&=&
-\frac{\varepsilon+p}{nT^2}
\left(
\nabla^\mu T
+
T\dot{u}^\mu
\right)\nn\\
&=&
-\frac{\varepsilon+p}{nT^2}\,
F_E^\mu
\eeqa

From the transformation between the Eckart and LL frames derived earlier, the heat-flow vector and diffusion current satisfy
\beqa
q_E^\mu
=
-\frac{\varepsilon+p}{n}
\nu_L^\mu.
\label{eq:current_relation}
\eeqa

\subsubsection{Transformation of the thermal conductivity}

Substituting Eqs.~(\ref{eq:Eckart_law}) and (\ref{eq:LL_law}) into Eq.~(\ref{eq:current_relation}) yields
\beqa
-\kappa_E F_E^\mu
&=&
-\frac{\varepsilon+p}{n}
\left(
-\kappa_LF_L^\mu
\right)
\nn\\
&=&
\frac{\varepsilon+p}{n}
\kappa_LF_L^\mu\nn\\
&=&
\frac{\varepsilon+p}{n}
\kappa_L
\left[
-\frac{\varepsilon+p}{nT^2}
F_E^\mu+\frac{\dot{u}^{\mu}}{T}
\right]???
\nn\\
&=&
-\kappa_L
\frac{(\varepsilon+p)^2}{n^2T^2}
F_E^\mu.
\eeqa
Since \(F_E^\mu\) is arbitrary, we obtain
\beqa
\kappa_E
=
\kappa_L
\frac{(\varepsilon+p)^2}{n^2T^2}+\mathcal{O}(2).
\eeqa
Therefore,
\beqa
\kappa_L
=
\kappa_E
\left(
\frac{nT}{\varepsilon+p}
\right)^2+\mathcal{O}(2).
\label{eq80}
\eeqa

This relation shows that the thermal conductivity in the two  choices of frames differs only by a multiplicative thermodynamic factor. The transport coefficients are generally defined as the ratio between the flux and its corresponding force i.e., flux$=-$(transport coefficient)$\times$force. In the Eckart's frame, when an observer rides with the particles i.e., $u^{\mu}\,||\,N^{\mu}$, the flux is the heat $q^{\mu}$ that flows relative to the particles. Whereas, in LL frame, when $u^{\mu}\,||\,T^{\mu\nu}$, the particle diffusion $\nu^{\mu}$ acts as the flux and there is flow of particles relative to the energy. For this reason, a {\it{temperature gradient $\nabla T$}} appears in the Eckart frame for heat conduction and {\it{chemical potential gradient $\nabla \mu$}} appears in the LL frame to signify the particle diffusion. The currents in LL frame and Eckart frame are physically different. Because the fluid has a finite enthalpy density $\varepsilon+P$, if particles are diffusing relative to the energy, they carry enthalpy with them. Consequently, the heat flux  emerge in Eckart  frame is directly proportional to the particle diffusion we observe in LL frame as of Eq.\eqref{eq65}. A diffusion of one baryon carries an average enthalpy $\frac{\varepsilon+P}{n}$ with it. Therefore, the flux we measure changes by a factor of $\frac{\varepsilon+P}{n}$. If the flux changes, the coefficient linking it to the gradient must change accordingly to compensate.


The shift is not a change in the fluid's intrinsic ability to transport conserved quantities, but rather a smart mathematical recalibration of the constitutive relations to ensure that all physical observables like the entropy production remain independent of the observer's frame. This shift depends exclusively on the equilibrium thermodynamic variables ($\varepsilon,\,P,\,n,\,T$), and is therefore a constant translation in the space of dissipative currents.

\subsection{Second-order coefficients in MIS theory}
\label{sec6E}

In the MIS formulation, the entropy current contains quadratic corrections in the dissipative fluxes. In the Eckart frame, it may be written as
\beqa
S_E^\mu
=
su_E^\mu
+
\frac{q_E^\mu}{T}
-
\frac{\beta_2^{(E)}}{2T}
q_{E\nu}q_E^\nu
u_E^\mu
+
{\cal O}({\rm diss}^3),
\eeqa
where \(\beta_2^{(E)}\) is associated with the relaxation of the heat-flow vector.

In the LL frame, the entropy current takes the form
\beqa
S_L^\mu
=
su_L^\mu
-
\frac{\mu}{T}\nu_L^\mu
-
\frac{\beta_2^{(L)}}{2T}
\nu_{L\nu}\nu_L^\nu
u_L^\mu
+
{\cal O}({\rm diss}^3).
\eeqa

Using
\beqa
u_L^\mu
=
u_E^\mu
+
\frac{q_E^\mu}{\varepsilon+p}
+
{\cal O}(2),
\eeqa
and
\beqa
\nu_L^\mu
=
-\frac{n}{\varepsilon+p}
q_E^\mu
+
{\cal O}(2),
\eeqa
one obtains
\beqa
S_L^\mu
&=&
su_E^\mu
+
\frac{s}{\varepsilon+p}
q_E^\mu
+
\frac{\mu n}{T(\varepsilon+p)}
q_E^\mu
\nn\\
&&
-
\frac{\beta_2^{(L)}}{2T}
\frac{n^2}{(\varepsilon+p)^2}
q_{E\nu}q_E^\nu
u_E^\mu
+
{\cal O}(3).
\eeqa
Using Eq.\eqref{eq4}, we obtain
\beqa
S_L^\mu
=
su_E^\mu
+
\frac{q_E^\mu}{T}
-
\frac{\beta_2^{(L)}}{2T}
\frac{n^2}{(\varepsilon+p)^2}
q_{E\nu}q_E^\nu
u_E^\mu
+
{\cal O}(3).
\eeqa

As the entropy production is same from both the frame, we compare entropy current of Eckart and LL frame to obtain
\beqa
\beta_2^{(L)}
\frac{n^2}{(\varepsilon+p)^2}
=
\beta_2^{(E)},
\eeqa
or
\beqa
\beta_2^{(L)}
=
\beta_2^{(E)}
\left(
\frac{\varepsilon+p}{n}
\right)^2.
\eeqa
The difference between the coefficients is therefore a purely thermodynamic quantity and is independent of the dissipative fluxes. Similar constant translations arise for the remaining second-order coefficients \(\beta_0\) and \(\beta_1\). These transformations ensure that physical observables remain invariant under a change of hydrodynamic frame when all transport coefficients are transformed consistently.

\section{Choice of $\mathcal{Q}^{\mu}$,  entropy production, and the order of relativistic hydrodynamics}
\label{sec5}
A primary objective of any theory of dissipative hydrodynamics is to describe the irreversible processes that drive a system toward thermodynamic equilibrium. The fundamental principle governing such processes is the second law of thermodynamics, which states that the entropy of an isolated system can never decrease. Consequently, entropy serves as a natural measure of irreversibility and provides a powerful criterion for constraining the dynamics of nonequilibrium systems.

In thermodynamic equilibrium, the system remains in a stationary state in which all macroscopic observables remain constant in time. Since no dissipative processes occur in equilibrium, there is no entropy production and the entropy of the system remains constant. By contrast, when the system is driven away from equilibrium, gradients of temperature, density, or velocity give rise to irreversible transport phenomena such as heat conduction, diffusion, and viscous flow. These processes inevitably generate entropy, causing the system to evolve toward a new equilibrium configuration.

The requirement of non-negative entropy production constitutes one of the most important constraints in relativistic dissipative hydrodynamics. In the covariant formulation, this requirement is expressed locally through the inequality $\partial_{\mu}S^{\mu}\geq 0$. Any acceptable hydrodynamic theory must therefore satisfy this condition identically.

The entropy current introduced in Eq.~\eqref{eq12},
\beqa
S^{\mu}
=
p\beta^{\mu}
-\alpha N^{\mu}
-\beta_{\nu}T^{\mu\nu}
-\mathcal{Q}^{\mu},
\eeqa
contains an additional contribution \(\mathcal{Q}^{\mu}\), which accounts for nonequilibrium corrections arising from dissipative processes. The precise form of \(\mathcal{Q}^{\mu}\) plays a crucial role in determining the structure of the hydrodynamic theory. In particular, it governs the entropy production rate and ultimately determines the constitutive equations satisfied by the dissipative fluxes.

The simplest possibility is to retain only terms that are linear in the dissipative quantities and neglect all higher-order contributions. Such an approximation corresponds to considering only first-order deviations from local thermodynamic equilibrium. Within Eckart's formulation, the nonequilibrium contribution to the entropy current is assumed to be proportional to the heat-flow vector and is chosen as
\beqa
\mathcal{Q}^{\mu}
=
\frac{q^{\mu}}{T}.
\label{eq91}
\eeqa
Substituting this expression into the general form of the entropy current yields
\beqa
S^{\mu}
=
p\beta^{\mu}
-\alpha N^{\mu}
-\beta_{\nu}T^{\mu\nu}
-\frac{q^{\mu}}{T}.
\eeqa

Since \(\mathcal{Q}^{\mu}\) contains only terms linear in the dissipative fluxes, the resulting theory belongs to the class of first-order relativistic hydrodynamic theories. In such theories, the dissipative fluxes are directly proportional to first-order gradients of the thermodynamic variables. Consequently, the bulk viscous pressure, heat flux, and shear stress tensor are all first-order quantities in the gradient expansion.

Although this approach provides a relativistic generalization of the classical theories of irreversible thermodynamics, it suffers from several fundamental shortcomings. Most notably, the resulting equations of motion are parabolic in nature and permit instantaneous propagation of disturbances. This implies that signals can propagate with arbitrarily large velocities, thereby violating the principle of causality. Moreover, the equilibrium state is found to be unstable under certain classes of small perturbations.

To overcome these difficulties, M\"uller and subsequently Israel and Stewart generalized the entropy current by allowing \(\mathcal{Q}^{\mu}\) to contain terms quadratic in the dissipative fluxes. Schematically,
\beqa
\mathcal{Q}^{\mu}
\sim
q^{2},
\quad
\Pi^{2},
\quad
\pi^{\alpha\beta}\pi_{\alpha\beta},
\quad
q^{\alpha}\Pi,
\quad \cdots
\eeqa
in addition to the linear terms. The inclusion of these second-order contributions leads to relaxation-type evolution equations for the dissipative currents and introduces finite relaxation times into the theory. As a consequence, the hydrodynamic equations become hyperbolic and disturbances propagate with finite speeds, restoring both causality and stability.

Therefore, the order of a relativistic hydrodynamic theory is determined by the highest-order dissipative contributions retained in the entropy current. Theories in which \(\mathcal{Q}^{\mu}\) contains only linear terms correspond to first-order formulations such as the Eckart and Landau--Lifshitz theories, whereas the inclusion of quadratic terms gives rise to second-order theories, well known as the M\"uller--Israel--Stewart framework. 

In the original Eckart formulation, a minimal, first-order ansatz was adopted as Eq.\eqref{eq91} to provide
\beqa
S^{\mu}=s u^{\mu}+\frac{q^{\mu}}{T}\,.
\label{eq93}
\eeqa
Here, the first term $s u^{\mu}$ represents the entropy convected with the fluid motion, while the second term accounts for the entropy flux due to heat conduction. Inserting this expression into the second law, i.e., $\pd_{\mu}S^{\mu}\ge 0$, using the first law of thermodynamics, and the conservation equations $\partial_{\mu}T^{\mu\nu}=0$ and $\partial_{\mu}N^{\mu}=0$, one obtains the local entropy production rate:
\beqa
T\partial_{\mu}S^{\mu}
= -\Pi \,\partial_{\mu}u^{\mu}
  -q^{\mu}\left[\frac{1}{T}\partial_{\mu}T+u^{\nu}\partial_{\nu}u_{\mu}\right]
  -\pi^{\mu\nu}\langle\partial_{\mu}u_{\nu}\rangle\,,
\label{eq0137}
\eeqa
where the angled brackets denote the symmetric, traceless, and transverse projection of a second-rank tensor defined as
\beqa
\langle\partial_{\mu}u_{\nu} \rangle
= \frac{1}{2}\Delta^{\alpha}_{\mu}\Delta^{\beta}_{\nu}
\left(\partial_{\alpha}u_{\beta}+\partial_{\beta}u_{\alpha}
-\frac{2}{3}\Delta_{\alpha\beta}\,\partial_{\gamma}u^{\gamma}\right)\,,
\label{eq0138}
\eeqa
To guarantee that the inequality of the second law, it is sufficient to assume linear, algebraic relations between the dissipative fluxes and their conjugate thermodynamic forces. The dissipative fluxes in first-order theory (NS) are thus expressed as:
\beqa
\Pi_{E} &=& -\zeta_{E} \,\partial_{\mu}u^{\mu}\,,\\
\pi^{\mu\nu}_{E} &=& -2\eta_{E} \,\langle\partial^{\mu}u^{\nu}\rangle\,,\\
q_{E}^{\mu} &=& -\kappa_{E} \left[\frac{1}{T}\partial^{\mu}T+u^{\nu}\partial_{\nu}u^{\mu}\right]\,.
\label{eq98}
\eeqa
The positive coefficients $\zeta$, $\eta$, and $\kappa$ are respectively the bulk viscosity, shear viscosity, and thermal conductivity. Substituting these into Eq.~\eqref{eq0137} casts the entropy production into a manifestly positive quadratic form:
\beqa
T\partial_{\mu}S^{\mu}
= \frac{\Pi^{2}}{\zeta T}
+\frac{q^{\mu}q_{\mu}}{\kappa T^{2}}
+\frac{\pi^{\mu\nu}\pi_{\mu\nu}}{2\eta T}
\ge 0\,.
\label{eq0140}
\eeqa
To identify the dissipative fluxes in the LL frame, we substitute the appropriate expressions for $T^{\mu\nu}$, and $N^{\mu}$ in terms of $\nu^{\mu}$ into Eq.\eqref{eq93}. Adapting the similar procedure as before, we identify the dissipative fluxes in LL frame as:
\beqa
\Pi_{L} &=& -\zeta_{L} \,\partial_{\mu}u^{\mu}\,,\\
\pi^{\mu\nu}_{L} &=& -2\eta_{L} \,\langle\partial^{\mu}u^{\nu}\rangle\,,\\
q^{\mu}_{L} &=& -\kappa_{L} \left[\frac{nT}{\varepsilon+P}\pd^\mu\alpha\right]\,.
\label{eq102}
\eeqa

To overcome these fundamental issues, M\"uller \cite{Muller:1967zza}, and subsequently Israel and Stewart \cite{Israel:1976tn,Stewart,Israel:1979wp}, generalized the theory to second order in the dissipative fluxes. Their fundamental aspect was to promote the dissipative fluxes ($\Pi$, $q^{\mu}$, $\pi^{\mu\nu}$) to independent dynamical variables, thereby introducing relaxation effects that restore causality. This is achieved by extending the entropy current to include terms quadratic in the dissipative fluxes. The general ansatz reads:
\beqa
S^{\mu}= s u^{\mu}
&+& \frac{q^{\mu}}{T}
- \left[\beta_0\Pi^2 - \beta_1 q_\nu q^\nu 
	+ \beta_2\pi_{\rho\sigma} \pi^{\rho\sigma}\right] \frac{u^\mu}{2T} + \left[\alpha_0\Pi\Delta^{\mu\nu} + \alpha_1\pi^{\mu\nu}\right]\frac{q_\nu}{T}~,
\label{eq0141}
\eeqa
where $\beta_{0}, \beta_{1}, \beta_{2}$ are the relaxation coefficients that quantify the deviation of the entropy density from its equilibrium value, and $\alpha_{0}, \alpha_{1}$ are the coupling coefficients that account for the interplay between heat flow and viscous stresses (thermo-viscous coupling).

Applying the second law $\partial_{\mu}S^{\mu}\ge 0$ to Eq.~\eqref{eq0141} and treating $\Pi$, $q^{\mu}$, and $\pi^{\mu\nu}$ as independent variables yields evolution equations of the relaxation type. In the Eckart frame,  these equations take the form \cite{Israel:1979wp}:
\beqa
\label{eq0142}
\Pi_{E} &=& -\frac{1}{3}\zeta_{E}\left[\partial_{\mu}u^{\mu}
+\beta^{E}_{0}D\Pi-{\alpha}^{E}_{0}\,\partial_{\mu}q^{\mu}\right]~, \\
\label{eq0143}
\pi^{\mu\nu}_{E} &=& -2\eta_{E} \Delta^{\mu\nu\alpha\beta}
\left[\partial_{\alpha}u_{\beta}
+\beta^{E}_{2}D\pi_{\alpha\beta}-{\alpha}_{1}^{E}\partial_{\alpha}q_{\beta}\right]~,\\
\label{eq0144}
q^{\mu}_{E} &=& -\kappa_{E} T\Delta^{\mu\nu}
\left[\frac{1}{T}\partial_\nu T +Du_{\nu}
+{\beta}^{E}_{1} D{q_\nu}
-{\alpha}^{E}_{0}\partial_\nu \Pi 
-{\alpha}^{E}_{1}\partial_{\lambda}\pi^{\lambda}_{\nu}\right]~,
\eeqa
where $D\equiv u^{\mu}\partial_{\mu}$ denotes the comoving (substantial) derivative.

In the LL frame, where the velocity is aligned with the energy flow ($q_L^\mu=0$), and the particle diffusion current $\nu^\mu$ becomes the relevant dissipative flux. The corresponding evolution equations for the remaining dissipative quantities are modified to
\beqa
\Pi_{L} &=&-\frac{1}{3}\zeta_{L}\left[\partial_{\mu}u^\mu 
+\beta^{L}_0 D \Pi-\alpha^{L}_0  \partial_{\mu}q^\mu \right]~,\nonumber\\   
\pi^{\mu\nu}_{L} &=& -2\eta_{L} \Delta^{\mu\nu\alpha\beta}
\left[\partial_{\alpha}u_{\beta}
+\beta^{L}_{2}D\pi_{\alpha\beta}-\alpha^{L}_{1}\partial_{\alpha}q_{\beta}\right]~,\nonumber\\
q^{\mu}_{L} &=& \kappa_{L} T\Delta^{\mu\nu}
\left[\frac{nT}{\varepsilon+P}(\partial_\nu \alpha )
-\beta^{L}_1 D{q_\nu}
+\alpha^{L}_0\partial_\nu \Pi 
+\alpha^{L}_1\partial_{\lambda}\pi^{\lambda}_{\nu} \right]~. 
\label{eq0145}
\eeqa
Crucially, the relaxation and coupling coefficients are not invariant under a change of the hydrodynamic frame. The Eckart-frame coefficients (${\alpha}^{E}_{0}, {\alpha}^{E}_{1}, {\beta}^{E}_{1}$) are related to their LL-frame counterparts ($\alpha^{L}_{0}, \alpha^{L}_{1}, \beta^{L}_{1}$) by the following constant shifts \cite{Israel:1979wp} as discussed in Sec.\ref{sec6E}.


The relaxation coefficients $\beta_0, \beta_1, \beta_2$ are directly related to the characteristic relaxation time scales of the dissipative processes \cite{Muronga:2003ta,Muronga:2001zk}:
\beqa
\tau_{\Pi}=\zeta \beta_0, \qquad
\tau_q=\kappa T\beta_1, \qquad
\tau_{\pi}=2\eta \beta_2~.
\label{eq0147}
\eeqa
Likewise, the coupling coefficients $\alpha_0$ and $\alpha_1$ determine the relaxation lengths that couple heat flux with bulk pressure and shear stress, respectively:
\beqa
l_{\Pi q}=\zeta \alpha_0,\qquad
l_{q\Pi}=\kappa T \alpha _0, \qquad
l_{q\pi}=\kappa T\alpha_1,\qquad 
l_{\pi q}=2\eta \alpha_1~.
\label{eq0148}
\eeqa
These length scales characterize the spatial range over which cross-coupling between different dissipative channels is significant.

Thus, the choice of $\mathcal{Q}^{\mu}$-whether it contains only linear terms (NS) or quadratic terms (MIS)-determines both the order of the theory and the precise definitions of the dissipative fluxes. The frame dependence of the transport coefficients is not a physical ambiguity but rather a reflection of the fact that different observers define `rest' and `dissipation' differently; physical observables such as the entropy production and the momentum flow remain invariant once the coefficients are transformed consistently according to the Sec.\ref{sec4}. The explicit expressions for these coefficients, evaluated from kinetic theory via thermodynamic integrals, are provided in Appendix~\ref{appA}.
\section{Sound Attenuation and Frame Invariance}
\label{sec6}
While the thermal conductivity undergoes a thermodynamic scaled transformation between the Eckart and Landau-Lifshitz frames, a physically observable quantity must remain independent of the hydrodynamic frame. To verify this explicitly, we examine the propagation and damping of longitudinal sound modes in a baryon-rich relativistic fluid by the linearization technique for the hydrodynamic variables.

\subsection{Linearized Hydrodynamic Equations}
Consider a homogeneous equilibrium state characterized by constant temperature $T_0$, baryon density $n_0$, pressure $P_0$, and energy density $\varepsilon_0$. Small perturbations around equilibrium are introduced as
\begin{equation}
\varepsilon=\varepsilon_0+\delta\varepsilon,
\qquad
n=n_0+\delta n,
\qquad
u^\mu=(1,\delta u,0,0),
\end{equation}

where $|\delta \varepsilon| \ll \varepsilon_0$, $|\delta n| \ll n_0$, and $|\delta u| \ll 1$. Restricting the analysis to longitudinal perturbations propagating along the $x$-direction, we assume plane-wave solutions of the form
\begin{equation}
\delta X(t,x)
=
\delta X_0
e^{i{(kx-\omega t)}}.
\end{equation}
The conservation laws are followed by the Eqs.\eqref{eq25} and \eqref{eq26} which  provides a set of coupled algebraic equations for the perturbations.
\subsubsection{Sound modes in the LL frame}
In the Landau-Lifshitz frame, the energy flux vanishes, and dissipation is carried by the particle diffusion current $\nu^\mu$. Linearizing the conservation equations yields
\beqa
\label{eq110}
0&=&-i\omega\,\delta\varepsilon
+
ik(\varepsilon_{0}+P_{0})\delta u,
\\
0&=&-i\omega(\varepsilon_{0}+P_{0})\delta u
+
ik\,\delta P
+
\left(
\frac{4}{3}\eta+\zeta
\right)k^2\delta u,\\
\label{eq112}
0&=&-i\omega\,\delta n
+
ik n_{0}\,\delta u
+
ik\delta\nu.
\eeqa
The diffusion current is related to fluctuations of the generalized thermodynamic force through
\begin{equation}
\delta\nu
=
-ik\,\kappa_L
\left(
\frac{n_{0}T_{0}}{\varepsilon_{0}+P_{0}}
\right)^2
\,
\delta
\left(
\frac{\mu}{T}
\right).
\end{equation}
The thermodynamic fluctuations can be expressed as
\begin{equation}
\label{eq114}
\delta P
=
\left(
\frac{\partial P}{\partial \varepsilon}
\right)_n
\delta\varepsilon
+
\left(
\frac{\partial P}{\partial n}
\right)_\varepsilon
\delta n
=
P_\varepsilon\,\delta\varepsilon
+
P_n\,\delta n,
\end{equation}
where
\begin{equation}
P_\varepsilon
=
\left(
\frac{\partial P}{\partial\varepsilon}
\right)_n,
\qquad
P_n
=
\left(
\frac{\partial P}{\partial n}
\right)_\varepsilon.
\end{equation}
and
\begin{equation}
\delta
\left(
\frac{\mu}{T}
\right)
=
A\,\delta\varepsilon
+
B\,\delta n,
\end{equation}
where
\begin{equation}
A=
\left(
\frac{\partial(\mu/T)}
{\partial\varepsilon}
\right)_n,
\qquad
B=
\left(
\frac{\partial(\mu/T)}
{\partial n}
\right)_\varepsilon .
\end{equation}
The diffusion current is then given by
\begin{equation}
\delta\nu
=-ik
\kappa_L
\left(
\frac{n_{0}T_{0}}{\varepsilon_{0}+P_{0}}
\right)^2
\left(
A\,\delta\varepsilon
+
B\,\delta n
\right).
\end{equation}
Using the above equations and considering the equilibrium enthalpy density, $h_{0}=\varepsilon_{0}+P_{0}$ into the set of equations i.e., Eqs. \eqref{eq110}-\eqref{eq112}, we obtain:
\beqa
0&=&-i\omega\,\delta\varepsilon
+
ikh_{0}\,\delta u,\\
0&=&ikP_\varepsilon\,\delta\varepsilon
+
ikP_n\,\delta n
+
\left[
-i\omega h_{0}
+
\left(
\frac{4}{3}\eta+\zeta
\right)k^2
\right]
\delta u,\\
0&=&-ik\kappa_L
\left(
\frac{n_{0}T_{}}{h_{0}}
\right)^2
A\,\delta\varepsilon
-
\left[
i\omega
+
ik\kappa_L
\left(
\frac{n_{0}T_{0}}{h_{0}}
\right)^2
B
\right]
\delta n
+
ikn_{0}\,\delta u\,.
\eeqa
Collecting the coefficients of the perturbations
\(
\delta\varepsilon
\),
\(
\delta n
\),
and
\(
\delta u
\),
the system can be written compactly as
\begin{equation}
M^{L}(\omega,k)
\begin{pmatrix}
\delta\varepsilon\\
\delta n\\
\delta u
\end{pmatrix}
=
0,
\label{matrixeq}
\end{equation}
where
\begin{equation}
M^{L}(\omega,k)
=
\begin{pmatrix}
-i\omega
&
0
&
ikh_{0}
\\[2mm]

ikP_\varepsilon
&
ikP_n
&
-i\omega h_{0}
+
\left(
\frac{4}{3}\eta+\zeta
\right)k^2
\\[2mm]
-ik\kappa_L
\left(
\frac{n_{0}T_{0}}{h_{0}}
\right)^2 A
&
-i\omega
-
ik\kappa_L
\left(
\frac{n_{0}T_{0}}{h_{0}}
\right)^2 B
&
ikn_{0}
\end{pmatrix}.
\label{matrixM}
\end{equation}
The hydrodynamic modes are obtained from the dispersion relation, which may be obtained by putting
\begin{equation}
\det [M^{L}(\omega,k)]=0.
\label{dispersion}
\end{equation}
Equation~(\ref{dispersion}) constitutes the dispersion relation for the longitudinal hydrodynamic modes. Since the determinant is cubic in $\omega$, three hydrodynamic branches emerge. In the long-wavelength limit $k\to \,0$, two correspond to propagating sound modes,
\begin{equation}
\omega^{L}_{\pm}
=
\pm c^{L}_s k
-i\Gamma^{L}_s k^2
+\mathcal{O}(k^3),
\end{equation}
while the third corresponds to the diffusive baryon mode,
\begin{equation}
\omega^{L}_D
=
-iD^{L}k^2
+\mathcal{O}(k^4).
\end{equation}
The sound velocity $c^{L}_s$, diffusion coefficient $D^{L}$, and sound attenuation coefficient $\Gamma^{L}_s$ are obtained by expanding Eq.~(\ref{dispersion}) order-by-order in the wave number $k$. In particular, $\Gamma^{L}_s$ contains contributions from both viscous dissipation and baryon diffusion and therefore provides a physically meaningful quantity for examining the frame dependence of the transport coefficients. Therefore, we may write
\begin{equation}
\Gamma^{L}_s
=
\Gamma^{L}_{\rm visc}
+
\Gamma^{L}_{\rm diff}.
\end{equation}
The viscous contribution is governed by the shear and bulk viscosities, while the diffusive contribution depends explicitly on the thermal conductivity $\kappa_L$ and the thermodynamic susceptibilities. The viscous contribution is given by
\begin{equation}
\Gamma^{L}_{\rm visc}
=
\frac{1}{2(\varepsilon_{0}+P_{0})}
\left(
\frac{4}{3}\eta+\zeta
\right),
\label{gammavisc}
\end{equation}
which represents the damping arising from shear and bulk viscous stresses. The diffusive contribution originates from the coupling between density fluctuations and the baryon diffusion current. 
the diffusive correction to the attenuation coefficient becomes
\begin{equation}
\Gamma_{\rm diff}
=
\frac{1}{2}
\left(
\frac{\partial P}
{\partial n}
\right)_\varepsilon
D^{L}
\left(
\frac{\partial n}
{\partial P}
\right)_{s/n}.
\label{gammadiff}
\end{equation}
Introducing the baryon diffusion coefficient
\begin{equation}
D^{L}
=
\kappa_L
\left(
\frac{n T}{\varepsilon_{0}+P_{0}}
\right)^2
\left[
\left(
\frac{\partial(\mu_B/T)}
{\partial n}
\right)_\varepsilon
\right]^{-1},
\label{diffusioncoef}
\end{equation}
Combining Eqs.~(\ref{gammavisc}) and (\ref{gammadiff}), the sound attenuation coefficient in the Landau-Lifshitz frame can be written as

\begin{equation}
\Gamma_s^{L}
=
\frac{1}{2(\varepsilon_{0}+P_{0})}
\left(
\frac{4}{3}\eta+\zeta
\right)
+
\frac{1}{2}
\left(
\frac{\partial P}
{\partial n}
\right)_\varepsilon
D^{L}
\left(
\frac{\partial n}
{\partial P}
\right)_{s/n}.
\label{GammaLL}
\end{equation}

Substituting Eq.~(\ref{diffusioncoef}) into Eq.~(\ref{GammaLL}) yields
\begin{equation}
\Gamma_s^{L}
=
\frac{1}{2(\varepsilon_{0}+P_{0})}
\left(
\frac{4}{3}\eta+\zeta
\right)
+
\frac{1}{2}
\left(
\frac{\partial P}
{\partial n}
\right)_\varepsilon
\kappa_L
\left(
\frac{n T}
{\varepsilon_{0}+P_{0}}
\right)^2
\left[
\left(
\frac{\partial(\mu_B/T)}
{\partial n}
\right)_\varepsilon
\right]^{-1}
\left(
\frac{\partial n}
{\partial P}
\right)_{s/n}.
\label{GammaLLfinal}
\end{equation}
\subsubsection{Sound Modes in the Eckart Frame}
The pressure fluctuation from Eq.\eqref{eq114} may be written as
\begin{equation}
\delta P
=
P_\varepsilon\,\delta\varepsilon
+
P_n\,\delta n,
\label{dP}
\end{equation}
Similarly, the temperature fluctuation is
\begin{equation}
\delta T
=
T_\varepsilon\,\delta\varepsilon
+
T_n\,\delta n,
\label{dT}
\end{equation}
where
\begin{equation}
T_\varepsilon
=
\left(
\frac{\partial T}
{\partial\varepsilon}
\right)_{n},
\qquad
T_n
=
\left(
\frac{\partial T}
{\partial n}
\right)_\varepsilon.
\end{equation}
Linearizing the heat flux equation in Eckart frame, we write
\begin{equation}
\delta q
=
-\kappa_E
\left[
ik\,\delta T
+
ik\,\frac{T}{h}\,\delta P
\right].
\label{dq}
\end{equation}
Substituting Eqs.~(\ref{dP}) and (\ref{dT}) into Eq.~(\ref{dq}) yields
\begin{equation}
\delta q
=
-ik\kappa_E
\left[
\left(
T_\varepsilon+\frac{T_{0}}{h_{0}}P_\varepsilon
\right)\delta\varepsilon
+
\left(
T_n+\frac{T_{}}{h_{0}}P_n
\right)\delta n
\right].
\label{eq137}
\end{equation}
Defining
\begin{equation}
A_E
=
T_\varepsilon
+
\frac{T_{0}}{h_{0}}P_\varepsilon,
\qquad
B_E
=
T_n
+
\frac{T_{0}}{h_{0}}P_n,
\end{equation}
The Eq.\eqref{eq137} becomes
\begin{equation}
\delta q
=
-ik\kappa_E
\left(
A_E\,\delta\varepsilon
+
B_E\,\delta n
\right).
\end{equation}
Finally, the linearized equations appear as:
\beqa
0&=&ik\,\delta P
-i\omega h_{0}\,\delta u
+
\left(
\frac{4}{3}\eta+\zeta
\right)k^2\delta u
-i\omega\,\delta q\,,\\
0&=&\Bigl[
ikP_\varepsilon
-i\omega k\kappa_EA_E
\Bigr]
\delta\varepsilon
+\Bigl[
ikP_n
-i\omega k\kappa_EB_E
\Bigr]
\delta n+
\Bigl[
-i\omega h_{0}
+
\left(
\frac{4}{3}\eta+\zeta
\right)k^2
\Bigr]
\delta u\,,\\
0&=&-i\omega\,\delta n
+
ikn_{0}\,\delta u\,.
\eeqa
In the matrix form:
\begin{equation}
M_E(\omega,k)
=
\begin{pmatrix}
-i\omega
&
0
&
ikh_{0}
\\[2mm]

ikP_\varepsilon
-\omega k\kappa_EA_E
&
ikP_n
-\omega k\kappa_EB_E
&
-i\omega h_{0}
+
\left(
\frac{4}{3}\eta+\zeta
\right)k^2
\\[2mm]

0
&
-i\omega
&
ikn_{0}
\end{pmatrix}.
\label{MEfinal}
\end{equation}
The dispersion relations are obtained by taking
\begin{equation}
\det [M_E(\omega,k)]=0.
\label{dispersionEckart}
\end{equation}
Following the method adapted in LL frame of reference, the viscous contribution may be found as:
\begin{equation}
\Gamma_{\rm visc}
=
\frac{1}{2(\varepsilon_{0}+P_{0})}
\left(
\frac{4}{3}\eta+\zeta
\right)
\label{GammaViscE}
\end{equation}
which is identical to the corresponding Landau--Lifshitz result. The dissipative contribution associated with thermal transport can be expressed as
\begin{equation}
\Gamma_{\rm diff}^{E}
=
\frac{1}{2}
\left(
\frac{\partial P}
{\partial n}
\right)_\varepsilon
D^{E}
\left(
\frac{\partial n}
{\partial P}
\right)_{s/n}
\label{GammaDiffE}
\end{equation}
where
\begin{equation}
D^{E}
=
\kappa_E
\left(
\frac{n T}
{\varepsilon_{0}+P_{0}}
\right)^2
\left[
\left(
\frac{\partial(\mu_B/T)}
{\partial n}
\right)_\varepsilon
\right]^{-1}
\label{DE}
\end{equation}
is the diffusion coefficient expressed in terms of the Eckart thermal conductivity. Now, combining Eqs.~(\ref{GammaViscE}) and (\ref{GammaDiffE}), the sound attenuation coefficient takes the form
\begin{equation}
\Gamma_s^{E}
=
\frac{1}{2(\varepsilon_{0}+P_{0})}
\left(
\frac{4}{3}\eta+\zeta
\right)
+
\frac{1}{2}
\left(
\frac{\partial P}
{\partial n}
\right)_\varepsilon
D^{E}
\left(
\frac{\partial n}
{\partial P}
\right)_{s/n}\,.
\label{GammaFinalE}
\end{equation}

Using the conductivity transformation derived previously,
\begin{equation}
\kappa_L
=
\kappa_E
\left(
\frac{n T}
{\varepsilon_{0}+P_{0}}
\right)^2,
\label{kappatransformE}
\end{equation}
one finds
\begin{equation}
D^{E}
=
\kappa_L
\left[
\left(
\frac{\partial(\mu_B/T)}
{\partial n}
\right)_\varepsilon
\right]^{-1}
\equiv D^{L}.
\end{equation}

Consequently,
\begin{equation}
\Gamma_{\rm diff}^{E}
=
\Gamma_{\rm diff}^{L},
\end{equation}
and therefore
\begin{equation}
\Gamma_s^{E}
=
\Gamma_s^{L}.
\label{GammaInvariant}
\end{equation}

This result shows that the sound attenuation coefficient is a frame-invariant observable. Although the thermal conductivity transforms thermodynamically between the Eckart and LL frames, the associated thermodynamic forces transform simultaneously so that the combination entering the sound-mode dispersion relation remains unchanged. The frame dependence is therefore encoded in the definition of the dissipative currents, while physical observables such as the sound velocity and attenuation rate remain invariant.



\section{Results and discussion}
\label{sec7}
The thermal conductivity is not a frame-invariant transport coefficient. Its numerical value depends on the choice of hydrodynamic frame used to define the fluid four-velocity. In particular, the conductivities in the Eckart and LL frames are related through Eq.\eqref{eq80}, which follows directly from the transformation connecting the corresponding dissipative currents. To quantify the resulting frame dependence, we consider a baryon-rich relativistic fluid described by a Boltzmann nucleon gas equation of state,
\begin{equation}
P(T,\mu_B)=
\frac{g m_N^2 T^2}{2\pi^2}
K_2\!\left(\frac{m_N}{T}\right)
\cosh\!\left(\frac{\mu_B}{T}\right),
\end{equation}
\begin{equation}
n(T,\mu_B)=
\frac{g m_N^2 T}{2\pi^2}
K_2\!\left(\frac{m_N}{T}\right)
\sinh\!\left(\frac{\mu_B}{T}\right),
\end{equation}
with the energy density determined from
\begin{equation}
\varepsilon=
T\left(\frac{\partial P}{\partial T}\right)_{\mu_B}
-P+\mu_B n.
\end{equation}
Using the Eq.\eqref{eq4}, one may rewrite 
\begin{equation}
\frac{\kappa_L}{\kappa_E}
=
\left(
\frac{n T}
{Ts+\mu_B n}
\right)^2.
\end{equation}
\begin{figure}[h]
	\centering
	\includegraphics[width=0.48 \textwidth]{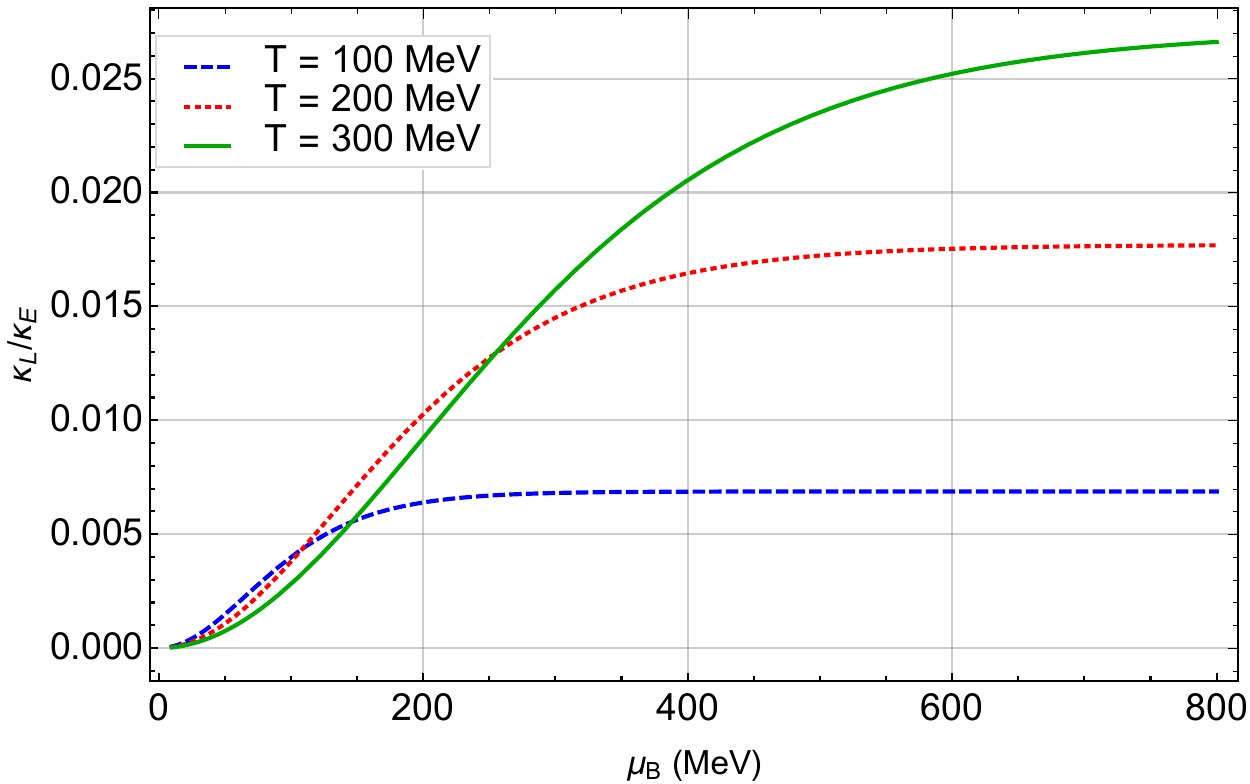}
	\includegraphics[width=0.48 \textwidth]{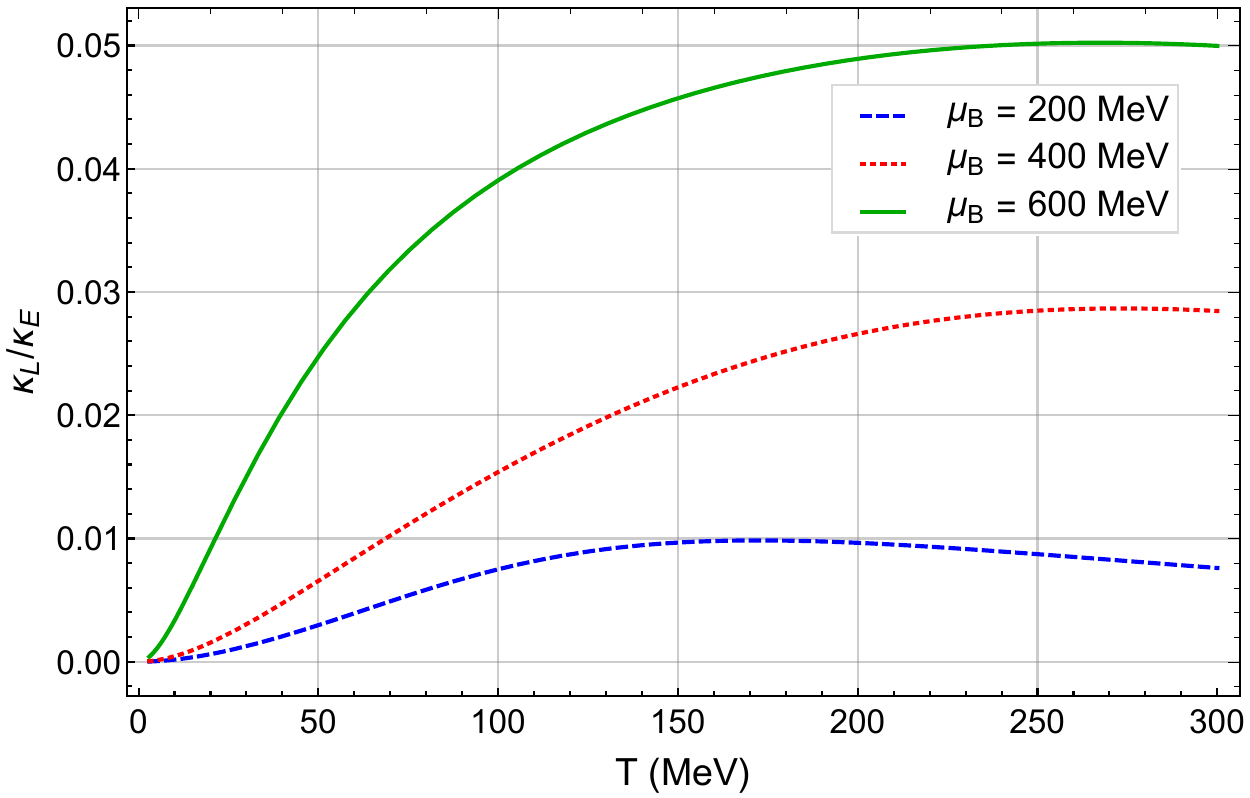}
	\caption{Left panel shows the ratio of the thermal conductivity of LL and Eckart frame with the variation of baryon chemical potential with different temperature. Right panel is the same ratio but with the variation of T with different chemical potential.}
	\label{fig1}
\end{figure}
The left panel of Fig.\ref{fig1} show the plot of $\kappa_{L}/\kappa_{E}$ as a function of baryon chemical potential with a set of temperatures. The plot demonstrates that the LL conductivity is smaller than its Eckart counterpart. The ratio increases with large range of $\mu_{B}$. With increasing temperature, the ratio also enhances. This behaviour is a direct consequence of the increasing enthalpy per baryon number, $(\varepsilon+P)/n$, which serves as the conversion factor linking heat transport and particle diffusion. In a baryon-rich environment, the contribution of the term $\mu_B n$ to the enthalpy density becomes substantial, thereby enhancing the disparity between $\kappa_{L}$ and $\kappa_{E}$.

It is important to emphasize, however, that this disparity does not signal a change in the underlying microscopic transport properties of the medium. The same collision dynamics and scattering processes govern both descriptions. Rather, the difference originates from the choice of the conserved currents. In the Eckart frame the dissipative nature is characterized by a heat current, whereas in the LL frame the same physical process is represented by the particle-diffusion current. The transformation of Eq.\eqref{eq80} therefore represents a reparametrization of the dissipative dynamics rather than a modification of the transport mechanism itself.

It is worth emphasizing that the conductivity coefficients themselves are not directly observable quantities. Their numerical values depend on the hydrodynamic frame used to decompose the conserved currents and the energy-momentum tensor. The relation between $\kappa_E$ and $\kappa_L$ therefore reflects a change of hydrodynamic variables rather than a modification of the microscopic dynamics. Quantities such as entropy production, diffusive attenuation rates, and the hydrodynamic evolution of the medium are unchanged once the corresponding dissipative currents and transport coefficients are transformed consistently. Accordingly, the difference between $\kappa_E$ and $\kappa_L$ represents a manifestation of hydrodynamic frame dependence and should not be interpreted as a physically measurable effect.

\section{Summary}
\label{sec8}
In this work, we have investigated the frame dependence of 
transport coefficients of relativistic dissipative hydrodynamics, with particular 
attention to the role of hydrodynamic frame choice and metric signature 
convention. By establishing the explicit boost 
transformation connecting the two frames at linear order in dissipation, we 
demonstrated that the equilibrium thermodynamic variables--energy density, 
pressure, and particle number density--remain frame-independent up to 
second order in the dissipative corrections. Interestingly, it is found that the transport coefficients between the Eckart and LL frames are related by a transformation equation as $\kappa_L = \kappa_E \left(\frac{nT}{\varepsilon + p}\right)^2$. It shows that the two conductivities differ by a purely thermodynamic 
factor governed by the enthalpy-to-particle-density ratio. Also, analogous constant translations are found for the second-order relaxation coefficients 
$\beta_0,\,\beta_1,\,\beta_2$ and the thermo-viscous coupling coefficients 
$\alpha_0,\,\alpha_1$. Evaluating this ratio for a baryon-rich relativistic 
fluid described by a Boltzmann nucleon gas equation of state, we find that 
$\kappa_L$ is suppressed relative to $\kappa_E$, with the 
disparity growing monotonically with both temperature and baryon chemical 
potential $\mu_B$, driven by the increasing enthalpy per baryon 
$(\varepsilon + P)/n$ in dense, hot matter. To verify frame invariance 
explicitly, we analyzed the propagation of longitudinal sound modes via 
linearized hydrodynamics in both frames and demonstrated that the sound 
attenuation coefficient $\Gamma_s$-- which receives contributions from 
both viscous dissipation and baryon diffusion-- is identical in the two 
frames, $\Gamma^E_s = \Gamma^L_s$, despite the thermal conductivity 
transforming non-trivially between them. This confirms that the frame 
dependence of $\kappa$ is not a physical ambiguity but an inherent 
consequence of how dissipative physics is partitioned between the heat-flow 
vector in the Eckart frame and the particle-diffusion current in the LL 
frame, with the two descriptions related by the enthalpy carried per 
diffusing baryon. Although transport coefficients may take different numerical values in different hydrodynamic frames, physical observables such as entropy production, sound velocity, and attenuation rates remain invariant when the corresponding frame transformations are applied consistently. Therefore, it is essential to specify the hydrodynamic frame explicitly when reporting transport coefficients extracted from heavy-ion collision data, lattice QCD calculations, or kinetic-theory studies.

\appendix
\section{Relaxation and coupling coefficients}
\label{appA}
In this Appendix, we summarize the explicit expressions for the relaxation parameters and coupling coefficients that enter the Israel--Stewart hydrodynamic equations~\cite{Israel:1979wp}. These coefficients are required for the numerical solution of the second-order dissipative hydrodynamic equations. They are given by
\beqa
\alpha_{0}&=&(D_{41}D_{20}-D_{31}D_{30})\Lambda \phi \Omega J_{21}J_{31}~,\\
\alpha_{1}&=& (J_{41}J_{42}-J_{31}J_{52}) \Lambda \phi J_{21}J_{31}~, \\
\beta_{0}&=& \frac{3\beta}{\phi^{2}\Omega^{2}}
[5J_{52}-\frac{3}{D_{20}}J_{31}(J_{31}J_{30}-J_{41}J_{20})
+J_{41}(J_{41}J_{10}-J_{31}J_{20})~,
\eeqa
and
\beqa
\beta_{1}&=& \frac{D_{41}}{\Lambda^{2}nmJ_{21}J_{31}}~, \\
\beta_{2}&=& \frac{\beta J_{52}}{2\phi^{2}}~.
\eeqa
Here, the auxiliary quantities appearing in the above expressions are defined as
\beqa
D_{rs}&=&J_{r+1,s}J_{r-1,s}-(J_{rs})^{2},\,\,\,\,
\phi=T(\varepsilon+P),\,\,\,\,
\psi=\frac{\varepsilon+p}{n_im_i}, \nn\\
\Lambda&=&1+5\left(\frac{\phi}{nm}\right)-\psi^{2},\,\,\,\,
\Omega=3\left(\frac{\pd\ln\phi}{\pd\ln n}\right)-5~.
\eeqa
In these relations, $r$ and $s$ denote integers, while $m_i$ and $n_i$ represent, respectively, the mass and number density of the (i)-th particle species. The quantities $\phi$, $\psi$, and $\Lambda$ are evaluated from the corresponding thermodynamic variables, namely the energy density $\varepsilon$, pressure $P$, and particle number densities $n_i$. For the numerical calculations presented in this work, the current masses of the up and down quarks are taken to be $10~\mathrm{MeV}$. $J_{rs}$ is defined as,
\beqa
J_{rs}=\frac{A_{0}}{(2s+1)!!}\int_{0}^{\infty} N \Delta Sinh^{2(s+1)}\mathcal{R} Cosh^{r-2s}\mathcal{R} d\mathcal{R}~,
\eeqa
and
\beqa
N=\frac{1}{exp(\beta Cosh \mathcal{R} -\alpha)-\varepsilon}~,
\eeqa
where, 
\beqa
\Delta=1+\varepsilon N,\,\,\,\, A_{0}=4\pi m_i^3~.
\label{eq14}
\eeqa

\bibliography{Framehydro}

\end{document}